# METASTABLE BOUND STATES OF THE QUASI–BIMAGNETOEXCITONS IN THE LOWEST LANDAU LEVELS APPROXIMATION

I.V. Podlesny[1], I.A. Zubac[1], Cam Ngoc Hoang[2], M.A. Liberman[3]

[1]*Institute of Applied Physics, Academic str. 5, Chisinau, MD–2028, Republic of Moldova*
[2]*Institute of Physics, Vietnam Academy of Science and Technology, 10 Dao Tan, Ba Dinh, Hanoi, Vietnam*
[3]*Nordic Institute for Theoretical Physics (NORDITA) KTH and Stockholm University, Roslagstullsbacken 23, Stockholm, SE–106 91, Sweden*

## Abstract

Four different spin structures of two electrons and of two holes situated on the lowest Landau levels (LLLs) are taken into account to investigate possible bound states of the two-dimensional magnetic biexciton formed of two magnetoexcitons with opposite wave vectors and antiparallel dipole moments. The singlet and triplet states of the spins of two electrons and of two holes separately, as well as of two para- and two ortho-magnetoexcitons are considered. The general expressions describing the binding energy of the bound states and the normalization conditions characterized by the parameter $\eta = \pm 1, \pm 1/2$ for the corresponding wave functions are derived. It is shown that for all four spin configurations the stable bound states in the LLLs approximation do not exist. The most favorable of the four considered spin configurations happened to be the triplet-triplet spin structure of two electrons and of two holes. In its frame a metastable bound state with activation barrier comparable with two ionization potentials of the magnetoexciton is revealed.

Keywords: semiconductor; exciton; magnetic field; interaction.

## 1. Introduction

Excitons and biexcitons in a strong magnetic field have been studied in [1–4] using similarity of hydrogen atoms and hydrogen molecule with excitons and biexcitons [5, 6]. Assuming nuclei (holes) to be sufficiently heavy to neglect their Landau quantization states, it was shown [3, 4] that in a strong magnetic field transition to the triplet metastable state $^3\Pi_u$ can explain of an alternative excitonic bound state, which was revealed in the form of *"X*-line" in the optical spectra in experimental studies of the stressed Ge crystal in a strong magnetic field [7]. According to [8, 9] the Coulomb exchange electron-hole interaction can lead to the formation of the para and ortho excitonic states, which influence the binding energy of the biexciton. It is expected that the triplet-triplet spin states of two-dimensional magnetic excitons can also form a metastable bound state of the magnetic biexciton similar to that proposed in [3, 4]. Like it was found for a hydrogen molecule [3, 4], one can also expect that in a strong magnetic field the binding energy of a biexciton can be quite large if the electron in one of the excitons occupies the excited Landau level.

For 2D magnetoexcitons only the spinless magnetoexciton-magnetoexciton interaction was considered and the influence of spin configurations was not taken into account until recently. Already in the papers [10–12] it was established that the magnetoexciton formed of electrons and holes lying on the lowest Landau levels (LLLs), being bound in the states with in-plane wave vectors $\vec{k}_\| = 0$, form an ideal, noninteracting gas. For the Landau gauge description 2D electrons and holes, which move with the resultant wave vector $\vec{k}_{\parallel} = 0$ on the surface of a layer in the perpendicular magnetic field, are subjected to the Landau quantization with the same gyration points. They have the quantum orbits with the same radii, which do not depend on the electron and hole masses, but only on the magnetic length. Such e-h pairs being bound by the direct Coulomb interaction form the bound states with wave vector $\vec{k}_{\parallel} = 0$, which look as the completely neutral compound particles. Therefore, magnetoexcitons with $\vec{k}_{\parallel} = 0$ cannot form a bound state. Only two magnetoexcitons with opposite nonzero wave vectors $\vec{k}$ and $-\vec{k}$ can form bound states with the resultant wave vector equal to zero. This possibility is investigated in the present paper.

In [13] it was shown that the interaction between two magnetoexcitons with wave vectors $\vec{k}_{\parallel} = 0$ can appear taking into account the influence of the excited Landau levels (ELLs) as well as the Rashba spin-orbit coupling (RSOC) generated by the perpendicular external electric field parallel to the magnetic field. Note, that all interactions and bound states investigated below have nothing to do with these supplementary influences and are based on the direct Coulomb interaction of electrons and holes with arbitrary masses situated on their LLLs. In the present paper we consider only the direct Coulomb electron-hole interaction accompanied by the successive kinematic exchange between two electrons or between two holes. We will assume that the influence of the Lorentz force and of the external magnetic field is much stronger than the Coulomb electron-hole interaction. The Lorentz force in turn gives rise to the Landau quantization of each particle, and to a localized electron cloud similar to some bell. The Landau quantization state is characterized by the cyclotron energy, by the radius of the orbit depending on the magnetic length only, as well as by the position of this orbit in the space. The gyration point for each orbit is determined by the wave vector of each particle. By this reason the position of magnetoexcitons depend on its resulting wave vector $\vec{k}$ with the vector $\vec{d}$ between two electron clouds, which is perpendicular to the wave vector $\vec{k}$ and its value is proportional to $|\vec{k}|$. Since the Coulomb interaction, which is responsible for the formation of the magnetoexciton, depends on the distance $d$ it depends on $k$, which means a strong dependence between the center of mass motion and the relative motion. This is a specific property of magnetoexcitons, unlike the Wannier-Mott excitons. The magnetoexciton can be viewed as electrical dipole with center of mass motion $k$ and the arm of the electric dipole $d$. If two centers of quantization are overlapping and have coinciding gyration points (the radii of these orbits do not depend on the mass), then they are exactly the same for the electron and for the hole and two overlapping clouds form a completely neutral particle. This property is known as a "hidden symmetry". The interaction between two such magnetoexcitons with $k=0$ equals exactly to zero.

We will consider the possibility to form the bound state of the molecular type — the magnetic biexciton of two magnetoexcitons. In the frame of the bound state, there are different distances between two magnetoexcitons, which are represented as two dipole moments. Because the wave vectors and the distances between two magnetoexcitons are interconnected due to the Heisenberg uncertainty relation, there is a large distances with small wave vectors and with a small dipole moments, as well as small distances with a large dipole moments. In the structure of molecule these two dipoles are changing their arms and changing the distance between each other. Therefore, they cannot be considered as two rigid dipoles. Nevertheless, to describe a bound state a localized wave function is necessary. For this purpose the wave function of relative motion is selected in a way to obtain the minimum energy of the Coulomb interaction in the frame of four localized particles forming the molecule. It is found that the better result can be obtained in the case when the relative motion function has the maximum on the ring in the momentum space and the minimum in the center.

We consider four particles with two spin orientations for electrons and for holes. The most interesting reciprocal orientations of the spins are the following. Spins of two electrons and of two holes are forming separately singlet and triplet combinations. For four particles we consider the spin structure with combination of the singlet structure for electrons and the singlet structure for holes, as well as the triplet structure for electrons and the triplet structure for holes. The mixed spin structures are forbidden due to the hidden symmetry. Another possibility is to combine relatively to each other the spin of electron and of the hole inside each magnetoexciton. In this case we can form para-exciton or ortho-exciton from one side, and para or ortho exciton from another side, which interact either as two para-magnetoexcitons, or two ortho-magnetoexcitons forming the molecule.

The possibility of two magnetoexcitons to form a bound state is examined considering the Feynman diagrams for possible interactions between particles. For a "stationary" case, when all particles exist before and after the Coulomb scattering, the Feynman diagrams contain of four lines: two solid lines representing the behavior of the electrons and two dashed lines representing the evolution of two holes in the frame of bound states. All matrix elements containing the Coulomb interaction integrals are calculated in the explicit analytical form. The obtained total energy of Coulomb interacting particles depends on the parameter $\alpha$ of the variational wave function and shows existence of the metastable state with a sufficiently high energy barrier of approximately two ionization potentials of free magnetoexcitons. The mean distance between the magnetoexcitons in the bound state is $R \sim l_0 \sqrt{\alpha}$, where $l_0$ is the magnetic length. The calculations with more cumbersome wave function corresponding to the bound state with resulting spin $S=0$ formed by two ortho-magnetoexcitons confirmed the earlier obtained results [14, 15] and allow to better understand the uniqueness of the triplet-triplet spin configuration studied earlier and its importance in similar biexciton spin structures in the absence of the magnetic field.

The paper is organized as follows. In the Section 2 the Hamiltonian of 2D electrons with spins and heavy holes with effective spins situated on their LLLs and interacting through the Coulomb interactions is introduced. The analytical and numerical solutions are presented in Section 3. The obtained results are analyzed in Section 4. We concluded in Section 5.

## 2. The Hamiltonian of the electron-hole system and the wave functions of the 2D quasi-bimagnetoexcitons

The description of 2D electrons and holes is considered in the Landau gauge, in which the charged particles have a free motion in one in-plane direction described by the plane waves with one-dimensional wave numbers $p$ and $q$ and undergo the quantized oscillations around the gyration points in the perpendicular direction. The quantum numbers $n_e = n_h = 0$ of the Landau quantized levels for electrons and holes will be omitted below. The creation and annihilation operators for electrons and for holes are denoted as $a^\dagger_{p,\sigma}, a_{p,\sigma}$ and $b^\dagger_{q,\sigma}, b_{q,\sigma}$, correspondingly. These operators have a supplementary spin label $\sigma = \pm 1/2$, which describes the spin projections of the conduction electrons and of the effective spin of the heavy holes.

We consider the Hamiltonian describing the Coulomb interaction of the 2D electrons and holes situated on their LLLs. For simplicity, we will neglect the electron-hole exchange interaction leading to the splitting of the ortho and para magnetoexciton energy levels. Nevertheless the spin structure of the para and ortho magnetoexcitons will be taken into account, but without the RSOC. Then the Hamiltonian can be written in the form

$$H^{LLL}_{Coul} = \frac{1}{2}\sum_{\vec{Q}} W(\vec{Q})[\hat{\rho}(\vec{Q})\hat{\rho}(-\vec{Q}) - \hat{N}_e - \hat{N}_h],$$

$$\hat{\rho}(\vec{Q}) = \hat{\rho}_e(\vec{Q}) - \hat{\rho}_h(\vec{Q}); W(\vec{Q}) = \frac{2\pi e^2}{\varepsilon_0 S |\vec{Q}|} e^{-\frac{Q^2 l_0^2}{2}}, \quad (1)$$

$$\hat{\rho}_e(\vec{Q}) = \sum_{t,\sigma} e^{iQ_y t l_0^2} a^\dagger_{t+\frac{Q_x}{2},\sigma} a_{t-\frac{Q_x}{2},\sigma}; \hat{N}_e = \hat{\rho}_e(0),$$

$$\hat{\rho}_h(\vec{Q}) = \sum_{t,\sigma} e^{-iQ_y t l_0^2} b^\dagger_{t+\frac{Q_x}{2},\sigma} b_{t-\frac{Q_x}{2},\sigma}; \hat{N}_h = \hat{\rho}_h(0),$$

where $\varepsilon_0$ is the dielectric constant, $S$ is the layer surface area. $\hat{\rho}_e(\vec{Q})$ and $\hat{\rho}_h(\vec{Q})$ are the electron and hole plasmon operators correspondingly.

The Hamiltonian (1) can be transcribed in the way

$$H_{Coul}^{LLL} = H_{e-e}^{LLL} + H_{h-h}^{LLL} + H_{e-h}^{LLL},$$

$$H_{e-e}^{LLL} = \frac{1}{2}\sum_{\vec{Q}}\sum_{p,q}\sum_{\sigma_1,\sigma_2} W(\vec{Q}) e^{-iQ_x Q_y l_0^2} e^{iQ_y(p-q)l_0^2} a_{p,\sigma_1}^\dagger a_{q,\sigma_2}^\dagger a_{q+Q_x,\sigma_2} a_{p-Q_x,\sigma_1},$$

$$H_{h-h}^{LLL} = \frac{1}{2}\sum_{\vec{Q}}\sum_{p,q}\sum_{\sigma_1,\sigma_2} W(\vec{Q}) e^{iQ_x Q_y l_0^2} e^{-iQ_y(p-q)l_0^2} b_{p,\sigma_1}^\dagger b_{q,\sigma_2}^\dagger b_{q+Q_x,\sigma_2} b_{p-Q_x,\sigma_1},$$

$$H_{e-h}^{LLL} = -\sum_{\vec{Q}}\sum_{p,q}\sum_{\sigma_1,\sigma_2} W(\vec{Q}) e^{iQ_y(p+q)l_0^2} a_{p,\sigma_1}^\dagger b_{q,\sigma_2}^\dagger b_{q+Q_x,\sigma_2} a_{p-Q_x,\sigma_1}. \quad (2)$$

The interaction coefficients depend only on the difference ($p-q$) for the electron-electron ($e-e$) and the hole-hole ($h-h$) interactions, and on the sum ($p+q$) for the electron-hole ($e-h$) interactions. The magnetoexciton creation operator introduced in [12] but with spin labels [15, 16] is

$$\hat{\psi}_{ex}^\dagger(\vec{k},\Sigma_e,\Sigma_h) = \frac{1}{\sqrt{N}}\sum_t e^{ik_y t l_0^2} a_{t+\frac{k_x}{2},\Sigma_e}^\dagger b_{-t+\frac{k_x}{2},\Sigma_h}^\dagger, \quad N = \frac{S}{2\pi l_0^2}; \quad l_0 = \sqrt{\frac{\hbar c}{eB}}. \quad (3)$$

Here $\vec{k}(k_x,k_y)$ is the vector of the center of mass in-plane motion, $t$ is the unidimensional vector of the relative $e-h$ motion with the function of the relative motion $e^{ik_y t l_0^2}$ in the momentum representation, which leads to the $\delta(y-k_y l_0^2)$ function of the relative motion in the real space representation. $N$ is the degree of the degeneracy of the Landau quantization levels, which is proportional to $S$. $B$ is the magnetic field strength, and $\Sigma_e, \Sigma_h = \pm 1/2$ are the spin quantum numbers.

The wave function of the magnetoexciton is

$$\left|\psi_{ex}\left(\vec{k},\Sigma_e,\Sigma_h\right)\right\rangle = \hat{\psi}_{ex}^\dagger\left(\vec{k},\Sigma_e,\Sigma_h\right)|0\rangle; \quad a_{t,\sigma}|0\rangle = b_{t,\sigma}|0\rangle = 0, \quad (4)$$

where $|0\rangle$ is the ground state of the system. The 2D magnetoexciton with wave vector $\vec{k} \neq 0$ has the form of an electric dipole with the arm $d = k l_0^2$ oriented perpendicularly to the wave vector $\vec{k}$. As it was shown in [10–12] two magnetoexcitons with wave vectors $\vec{k} = 0$ are similar to the neutral compound of particles, have no the dipole moments and do not interact through the Coulomb forces. On the contrary, two magnetoexcitons with nonzero wave vectors $\vec{k}_1$ and $\vec{k}_2$ do interact opening the possibility to form bimagnetoexcitons. The wave function of two magnetoexcitons with quantum numbers $\left|\vec{k},\Sigma_{e,1},\Sigma_{h,1}\right\rangle$ and $\left|-\vec{k},\Sigma_{e,2},\Sigma_{h,2}\right\rangle$ can be written as

$$\left|\psi_{ex,ex}\left(\vec{k},\Sigma_{e,1},\Sigma_{h,1};-\vec{k},\Sigma_{e,2},\Sigma_{h,2}\right)\right\rangle = \frac{1}{N}\sum_{t,s} e^{ik_y(t-s)l_0^2} a_{t+\frac{k_x}{2},\Sigma_{e,1}}^\dagger a_{s-\frac{k_x}{2},\Sigma_{e,2}}^\dagger b_{-s-\frac{k_x}{2},\Sigma_{h,2}}^\dagger b_{-t+\frac{k_x}{2},\Sigma_{h,1}}^\dagger |0\rangle. \quad (5)$$

The wave function of the quasi-bimagnetoexciton with wave vector $\vec{k}=0$ as a bound state of two magnetoexcitons with wave vectors $\vec{k}$ and $-\vec{k}$ and spin quantum numbers $\Sigma_{e,1}, \Sigma_{h,1}, \Sigma_{e,2}, \Sigma_{h,2}$ can be constructed as a superposition of the wave functions (5) introducing the wave function $\varphi_n(\vec{k})$ of the relative motion, which can play the role of the variational function determining the minimal energy of the bimagnetoexciton, as well as the density $|\varphi_n(\vec{k})|^2$ of the magnetoexcitons taking part in the formation of the bound state. In [8, 9] it was shown that the spin configurations of the bound states depend essentially on the ratio between the ortho-para exciton splitting and the binding energy of the biexciton. In a strong magnetic field these values are unknown for the magnetoexciton formation and one of the purpose is to determine one of them.

We will consider four different spin structures of the bound states. First of all we will construct the symmetric and antisymmetric superpositions of two electron spin states and two hole effective spin states in the form

$$\frac{1}{\sqrt{2}} \sum_{\Sigma_e = \pm 1/2} (\eta_e)^{\Sigma_e + 1/2} a^\dagger_{p,\Sigma_e} a^\dagger_{q,-\Sigma_e}; \quad \frac{1}{\sqrt{2}} \sum_{\Sigma_h = \pm 1/2} (\eta_h)^{\Sigma_h + 1/2} b^\dagger_{p,\Sigma_h} b^\dagger_{q,-\Sigma_h}; \quad \eta_e = \pm 1; \quad \eta_h = \pm 1, \quad (6)$$

In more general case we can take four different combinations of the bound states of the type

$$|\psi_{bimex}(0, \eta_e, \eta_h, \varphi_n)\rangle = \frac{1}{2N^{3/2}} \sum_{\Sigma_e = \pm 1/2} (\eta_e)^{\Sigma_e + 1/2} \sum_{\Sigma_h = \pm 1/2} (\eta_h)^{\Sigma_h + 1/2} \sum_{\vec{k}} \varphi_n(\vec{k})$$
$$\times \sum_{t,s} e^{ik_y(t-s)l_0^2} a^\dagger_{t+\frac{k_x}{2}, \Sigma_e} a^\dagger_{s-\frac{k_x}{2}, -\Sigma_e} b^\dagger_{-s-\frac{k_x}{2}, -\Sigma_h} b^\dagger_{-t+\frac{k_x}{2}, \Sigma_h} |0\rangle. \quad (7)$$

Their normalization integrals are

$$\langle \psi_{bimex}(0, \eta_e, \eta_h, \varphi_n) | \psi_{bimex}(0, \eta_e, \eta_h, \varphi_n) \rangle = 1 + \eta_e \eta_h - (\eta_e + \eta_h) L_n(\alpha);$$
$$L_n(\alpha) = \frac{1}{N^2} \sum_{\vec{k}} \sum_{\vec{\varkappa}} \varphi_n^*(\vec{\varkappa}) \varphi_n(\vec{k}) e^{i(k_y \varkappa_x - k_x \varkappa_y) l_0^2}, \quad (8)$$

where $\alpha$ is the variational parameter. One can see that in the case $\eta_e = -\eta_h$ the normalization integrals vanish and remain the unique possibilities: $\eta_e = \eta_h = \eta = \pm 1$.

Below we will suppose that both pairs of spins $(\Sigma_{e,1}, \Sigma_{e,2})$ and $(\Sigma_{h,1}, \Sigma_{h,2})$ are simultaneously in the states with the same $\eta_e = \eta_h = \eta = \pm 1$. The bimagnetoexciton wave functions in these conditions are

$$|\psi_{bimex}(0, \eta, \varphi_n)\rangle = \frac{1}{2N^{3/2}} \sum_{\Sigma_e, \Sigma_h} (\eta)^{\Sigma_e + \Sigma_h + 1} \sum_{\vec{k}} \varphi_n(\vec{k}) \sum_{t,s} e^{ik_y(t-s)l_0^2} a^\dagger_{t+\frac{k_x}{2}, \Sigma_e} a^\dagger_{s-\frac{k_x}{2}, -\Sigma_e} b^\dagger_{-s-\frac{k_x}{2}, -\Sigma_h} b^\dagger_{-t+\frac{k_x}{2}, \Sigma_h} |0\rangle. \quad (9)$$

Due to the hidden symmetry in the system related with the same radii of the Landau quantization orbits for electrons and for holes, which depend only on the magnetic length $l_0$ and do not depend on the effective masses $m_e$ and $m_h$, their normalization integrals are

$$\langle\psi_{bimex}(0,\eta,\varphi_n)|\psi_{bimex}(0,\eta,\varphi_n)\rangle = 2(1-\eta L_n(\alpha)). \tag{10}$$

Side by side with correlation of spins in the frame of two electrons and of two holes one can consider the correlations of spins in the frame of each electron-hole pair forming the magnetoexciton.

The wave function of the para magnetoexciton looks as

$$|\hat{\psi}_{ex,p}(\vec{k})\rangle = \frac{1}{\sqrt{2N}} \sum_{t,\sigma=\pm 1/2} e^{ik_y t l_0^2} a^\dagger_{t+\frac{k_x}{2},\sigma} b^\dagger_{-t+\frac{k_x}{2},-\sigma} |0\rangle. \tag{11}$$

It corresponds to the resultant spin of the e-h pair $S=0$ with the projection $S_z=0$.

There are three wave functions $\psi_{ex,or}(\vec{k},S,S_z)$ of the ortho magnetoexciton with resultant spin $S=1$ and its projections $S_z=0,\pm 1$

$$\psi_{ex,or}(\vec{k},1,0) = \frac{1}{\sqrt{2N}} \sum_{t,\sigma=\pm 1/2} e^{ik_y t l_0^2} (-1)^{\sigma+1/2} a^\dagger_{t+\frac{k_x}{2},\sigma} b^\dagger_{-t+\frac{k_x}{2},-\sigma} |0\rangle,$$

$$\psi_{ex,or}(\vec{k},1,\pm 1) = \frac{1}{\sqrt{N}} \sum_{t,\sigma=\pm 1/2} e^{ik_y t l_0^2} a^\dagger_{t+\frac{k_x}{2},\downarrow} b^\dagger_{-t+\frac{k_x}{2},\downarrow} |0\rangle. \tag{12}$$

The molecular states formed by two bound para magnetoexcitons can be described by the functions

$$\psi^{p,p}_{bimex}(0,\varphi_n) = \frac{1}{2N^{3/2}} \sum_{\vec{k}} \varphi_n(\vec{k}) \sum_{\sigma_1,\sigma_2=\pm 1/2} \sum_{s,t} e^{ik_y(t-s)l_0^2} a^\dagger_{t+\frac{k_x}{2},\sigma_1} a^\dagger_{s-\frac{k_x}{2},\sigma_2} b^\dagger_{-s-\frac{k_x}{2},-\sigma_2} b^\dagger_{-t+\frac{k_x}{2},-\sigma_1} |0\rangle. \tag{13}$$

They are characterized by the resultant spin ($S$) of four bound particles equal to zero ($S=0$). Their normalization integrals are

$$\langle\psi^{p,p}_{bimex}(0,\varphi_n)|\psi^{p,p}_{bimex}(0,\varphi_n)\rangle = 2 - L_n(\alpha). \tag{14}$$

Following [8, 9, 17, **Error! Reference source not found.**] two ortho magnetoexcitons forming the bound states with the resultant spin $S=0$ may be constructed in the form of the invariant including all three wave functions of both ortho-magnetoexcitons in the form

$$|\psi^{or,or}_{bimex}(0,\varphi_n)\rangle = \frac{1}{\sqrt{3N}} \sum_{\vec{k}} \varphi_n(\vec{k}) [\hat{\psi}^\dagger_{ex,or}(\vec{k},1,0)\hat{\psi}^\dagger_{ex,or}(-\vec{k},1,0) +$$

$$\hat{\psi}^\dagger_{ex,or}(\vec{k},1,1)\hat{\psi}^\dagger_{ex,or}(-\vec{k},1,-1) + \hat{\psi}^\dagger_{ex,or}(\vec{k},1,-1)\hat{\psi}^\dagger_{ex,or}(\vec{k},1,1)|0\rangle =$$

$$\frac{1}{\sqrt{3N^3}} \sum_{\vec{k}} \sum_{t,s} \varphi_n(\vec{k}) e^{ik_y(t-s)l_0^2} \left\{ \frac{1}{2} \left[ a^\dagger_{t+\frac{k_x}{2},\uparrow} a^\dagger_{s-\frac{k_x}{2},\uparrow} b^\dagger_{-s-\frac{k_x}{2},\downarrow} b^\dagger_{-t+\frac{k_x}{2},\downarrow} + \right. \right.$$

$$a^\dagger_{t+\frac{k_x}{2},\downarrow} a^\dagger_{s-\frac{k_x}{2},\downarrow} b^\dagger_{-s-\frac{k_x}{2},\uparrow} b^\dagger_{-t+\frac{k_x}{2},\uparrow} - a^\dagger_{t+\frac{k_x}{2},\uparrow} a^\dagger_{s-\frac{k_x}{2},\downarrow} b^\dagger_{-s-\frac{k_x}{2},\uparrow} b^\dagger_{-t+\frac{k_x}{2},\downarrow} -$$

$$\left. a^\dagger_{t+\frac{k_x}{2},\downarrow} a^\dagger_{s-\frac{k_x}{2},\uparrow} b^\dagger_{-s-\frac{k_x}{2},\downarrow} b^\dagger_{-t+\frac{k_x}{2},\uparrow} \right] +$$

$$\left. a^\dagger_{t+\frac{k_x}{2},\uparrow} a^\dagger_{s-\frac{k_x}{2},\downarrow} b^\dagger_{-s-\frac{k_x}{2},\downarrow} b^\dagger_{-t+\frac{k_x}{2},\uparrow} + a^\dagger_{t+\frac{k_x}{2},\downarrow} a^\dagger_{s-\frac{k_x}{2},\uparrow} b^\dagger_{-s-\frac{k_x}{2},\uparrow} b^\dagger_{-t+\frac{k_x}{2},\downarrow} \right\} |0\rangle. \tag{15}$$

Their normalization integrals are

$$\langle \psi_{bimex}^{or,or}(0,\varphi_n) | \psi_{bimex}^{or,or}(0,\varphi_n) \rangle = 2 + L_n(\alpha). \quad (16)$$

Taking into account the expressions for normalization integrals (10), (14) and (16), the bound states for four spin structures can be written in the universal form

$$\langle \psi_{bimex}(0,\eta,\varphi_n) | \psi_{bimex}(0,\eta,\varphi_n) \rangle = 2(1 - \eta L_n(\alpha)), \ \eta = \pm 1;$$

$$\langle \psi_{bimex}^{p,p}(0,\varphi_n) | \psi_{bimex}^{p,p}(0,\varphi_n) \rangle = 2(1 - \eta L_n(\alpha)), \ \eta = \frac{1}{2}; \quad (17)$$

$$\langle \psi_{bimex}^{or,or}(0,\varphi_n) | \psi_{bimex}^{or,or}(0,\varphi_n) \rangle = 2(1 - \eta L_n(\alpha)), \ \eta = -\frac{1}{2}.$$

All of them have the expression

$$\langle \psi_{bimex}(0,\eta,\varphi_n) | \psi_{bimex}(0,\eta,\varphi_n) \rangle = 2(1 - \eta L_n(\alpha)) \text{ with } \eta = \pm 1, \ \eta = \pm 1/2. \quad (18)$$

The chosen variational wave functions of the relative motion in the momentum and in the real space representations $\varphi_n(\vec{k})$ and $\psi_n(\vec{r})$, their normalization conditions and the main parameters are

$$\varphi_0(x) = (4\alpha)^{1/2} e^{-\alpha x^2}; \ \varphi_2(x) = (8\alpha^3)^{1/2} x^2 e^{-\alpha x^2}; \ x = kl_0, \ \frac{1}{N}\sum_{\vec{k}} |\varphi_n(\vec{k})|^2 = \int_0^\infty x dx |\varphi_n(x)|^2 = 1,$$

$$\psi_n(\vec{r}) = \int \varphi_n(\vec{k}) e^{i\vec{k}\vec{r}} d^2\vec{k} = \int_0^\infty x dx \varphi_n(x) J_0(x \cdot \frac{r}{l_0}), \ \psi_0(r) \sim e^{-\frac{r^2}{4\alpha l_0^2}}; \ \psi_2(r) \sim \left(1 - \frac{r^2}{4\alpha l_0^2}\right) e^{-\frac{r^2}{4\alpha l_0^2}}, \quad (19)$$

where $J_0(z)$ is the Bessel function of the zeroth order.

The selected trial wave functions depend only on the modulus $k = |\vec{k}|$. $\varphi_0(x)$ has the maximum at the point $x = 0$, the mean value $\overline{x^2} = 1/(2\alpha)$, the radius of the quantum state $\psi_0(\vec{r})$ equals to $a = 2l_0\sqrt{\alpha}$. The function $\varphi_2(\vec{k})$ has the maximum on the 2D ring with the radius $k_r = 1/(l_0\sqrt{\alpha})$. In the real space the function $\psi_2(r)$ has the maximum at the point $r_0 = 0$, the positive values up till the point $r_1 = a$, where it changes sign and achieves the minimum at the point $r_2 = l_0\sqrt{8\alpha}$. Its absolute value at the minimum is much smaller than it is in the maximum.

Calculating the overlapping integrals $L_n(\alpha)$, we obtain

$$L_n(\alpha) = \int_0^\infty x dx \int_0^\infty y dy \varphi^*(x) \varphi(y) J_0(xy); L_0(\alpha) = \frac{1}{\alpha + \frac{1}{4\alpha}}; L_2(\alpha) = \frac{2\alpha^2 - \frac{1}{2}}{\left(\alpha + \frac{1}{4\alpha}\right)^3}. \quad (20)$$

Fig. 1 shows the normalization integrals $(1 - \eta L_n(\alpha))$. As one can see the factor $(1 - L_0(\alpha))$ vanishes at the point $\alpha = 1/2$, which leads to a singularity of the inverse function. The inverse normalization integral $(1 - L_2(\alpha))^{-1}$ is regular at any values of $\alpha$.

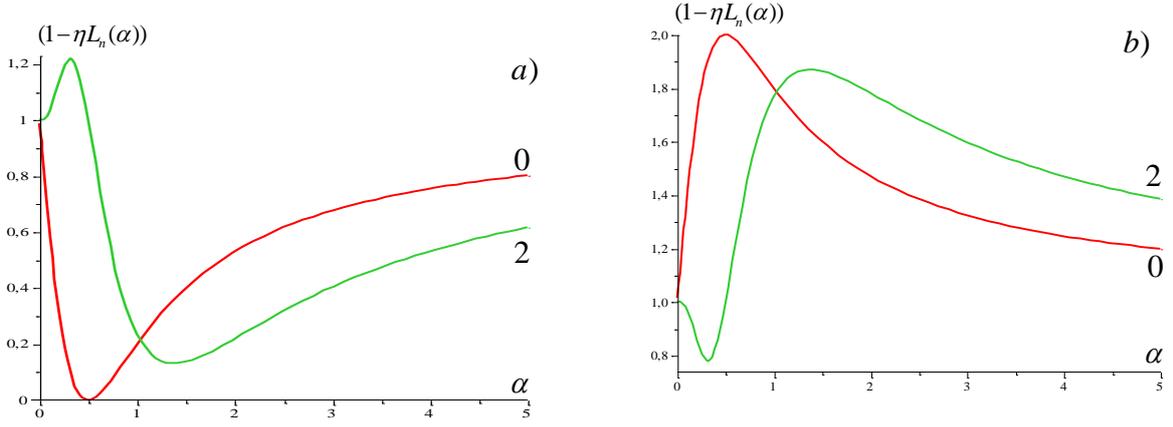

**Fig. 1.** The normalization integrals with $n = 0, 2$ in dependence of the parameter $\alpha$ in the cases: a) $\eta = 1$ and b) $\eta = -1$.

It should be mentioned that the case $\eta_e = \eta_h = \eta = \pm 1$ is the only possibility, because in the opposite case $\eta_e = -\eta_h$ the wave function of the type (9) and its normalization integrals vanish.

### 3. Binding energies of the lowest states of 2D bimagnetoexcitons

The expectation values of the Hamiltonian (2) averaged with the wave function (9) characterized by the wave vector $\vec{k} = 0$, values of $\eta = \pm 1, \pm 1/2$, and by the trial wave functions $\varphi_n(\vec{k})$ equal to

$$E_{bimex}(0, \eta, \varphi_n) = \frac{\langle \psi_{bimex}(0, \eta, \varphi_n) | H_{Coul}^{LLL} | \psi_{bimex}(0, \eta, \varphi_n) \rangle}{\langle \psi_{bimex}(0, \eta, \varphi_n) | \psi_{bimex}(0, \eta, \varphi_n) \rangle}. \tag{21}$$

Fig. 2 shows the Feynman diagrams describing the direct Coulomb interactions between electrons and holes accompanied with successive kinematic exchanges of homogeneous particles.

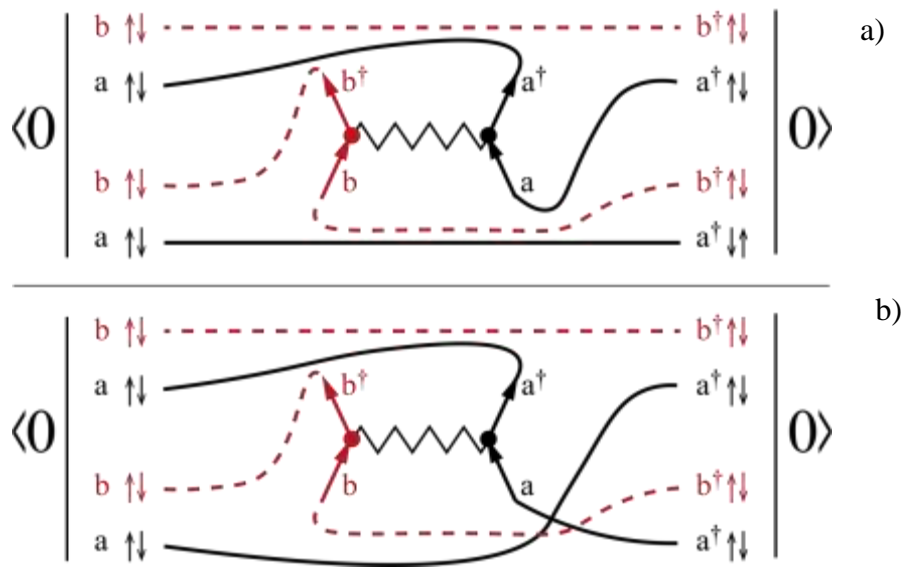

**Fig. 2.** The Feynman diagrams describing the direct Coulomb electron-hole interaction in the frame of the metastable bound state of two magnetoexcitons: a) with an intersection, and b) accompanied by the successive kinematic exchange between two electrons [17, 19].

Here we deal with the Feynman diagrams with participation of two pairs, rather than with one pair. Since two pairs of the particles take part in the Coulomb interaction, alongside with the direct interaction, the exchange interaction can take place. Therefore there can exist not only Coulomb direct dynamical interaction of two particles, but subsequently it can be accompanied by kinematic exchange interaction of two homogeneous particles. There is the Coulomb direct interaction between electrons and electrons, between holes and holes, and between electrons and holes accompanied by the exchange interaction of homogeneous particles — either by two electrons, or by two holes. The case of zero intersections in the diagrams and one intersection for electrons is shown in Fig. 2a and Fig. 2b, correspondingly. The case of double intersections with holes participation is also possible and will be considered below. The intrinsic exchange electron-hole interaction in the frame of one e-h pair is not taken into account. This means that after the Coulomb scattering process the electron is transformed into the hole and the hole into the electron. In these conditions the Coulomb charge-charge interaction is transformed into the dipole-dipole interaction, which is much smaller than the exciton binding energy.

The terms described by these diagrams as well as the similar one describing the electron-electron and the hole-hole Coulomb interactions are gathered in two groups introducing the spin structure index $\eta$ as it is shown in the scheme below

$$\langle \psi_{bimex}(0,\eta,\varphi_n) | H_{Coul}^{LLL} | \psi_{bimex}(0,\eta,\varphi_n) \rangle = \left\langle \begin{array}{c} \text{Feynman diagrams} \\ \text{without or with two} \\ \text{intersections} \end{array} \right\rangle + \eta \left\langle \begin{array}{c} \text{Feynman diagrams} \\ \text{with one} \\ \text{intersection} \end{array} \right\rangle,$$

$$\langle \psi_{bimex}(0,\eta) | \psi_{bimex}(0,\eta) \rangle = 2(1 - \eta L_n(\alpha)) \qquad (22)$$

with $\eta = 1$ describing the triplet-triplet spin structures of 2e+2h, $\eta = -1$ for singlet-singlet spin structures of 2e+2h; $\eta = 1/2, -1/2$ for ortho-ortho and para-para magnetoexcitons, correspondingly. In this way four different spin structures of the molecule turn out to be successfully represented in a unified manner (22). Unlike our previous work [14], here we consider four instead of two spin structures.

The average values of the partial Hamiltonians $H_{e-e}^{LLL}$ and $H_{h-h}^{LLL}$ calculated with the functions (9) can be expressed by

$$\langle \psi_{bimex}(0,\eta,\varphi_n) | H_{e-e}^{LLL} | \psi_{bimex}(0,\eta,\varphi_n) \rangle = \langle \psi_{bimex}(0,\eta,\varphi_n) | H_{h-h}^{LLL} | \psi_{bimex}(0,\eta,\varphi_n) \rangle =$$
$$= \frac{2}{N} \sum_{\vec{Q}} \sum_{\vec{\varkappa}} W(\vec{Q}) \varphi_n^*(\vec{\varkappa}) \varphi_n(\vec{Q}-\vec{\varkappa}) e^{i(\varkappa_x Q_y - \varkappa_y Q_x) l_0^2} - \qquad (23)$$
$$- \frac{2\eta}{N^2} \sum_{\vec{Q}} \sum_{\vec{\varkappa}} \sum_{\vec{k}} W(\vec{Q}) \varphi_n^*(\vec{\varkappa}) \varphi_n(\vec{k}) \exp\left\{ i l_0^2 \left[ (Q_x k_y - Q_y k_x) + (\varkappa_x Q_y - \varkappa_y Q_x) + (k_x \varkappa_y - k_y \varkappa_x) \right] \right\}.$$

Using the polar coordinates we can write

$$\vec{k} = (k_x, k_y) = k(\cos\varphi, \sin\varphi); \; \vec{k}l_0 = \vec{z}, \; kl_0 = z;$$
$$\vec{Q} = (Q_x, Q_y) = Q(\cos\theta, \sin\theta); \; \vec{Q}l_0 = \vec{y}; \; Ql_0 = y;$$
$$\vec{\varkappa} = (\varkappa_x, \varkappa_y) = \varkappa(\cos\psi, \sin\psi); \; \vec{\varkappa}d_0 = \vec{x}; \; \varkappa d_0 = x;$$
$$(Q_x k_y - Q_y k_x)l_0^2 = Qkl_0^2(\sin\varphi\cos\theta - \cos\varphi\sin\theta) = yz\sin(\varphi-\theta) = z_1\sin t_1;$$
$$(\varkappa_x Q_y - \varkappa_y Q_x)l_0^2 = Q\varkappa d_0^2(\sin\theta\cos\psi - \cos\theta\sin\psi) = xy\sin(\theta-\psi) = z_2\sin t_2;$$
$$(k_x \varkappa_y - k_y \varkappa_x)l_0^2 = k\varkappa d_0^2(\sin\psi\cos\varphi - \cos\psi\sin\varphi) = xz\sin(\psi-\varphi) = z_3\sin t_3,$$
(24)

where we introduced denotations

$$t_1 = \varphi - \theta; \; t_2 = \theta - \psi; \; t_3 = \psi - \varphi; \; z_1 = yz; \; z_2 = xy; \; z_3 = xz. \tag{25}$$

In the polar coordinates the expression (23) becomes

$$\langle \psi_{bimex}(0, \eta, \varphi_n) | H_{j-j}^{LLL} | \psi_{bimex}(0, \eta, \varphi_n) \rangle = \frac{2}{N} \sum_{\vec{Q}} \sum_{\vec{\varkappa}} W(\vec{Q}) \varphi_n^*(\vec{\varkappa}) \varphi_n(\vec{Q} - \vec{\varkappa}) e^{iz_2 \sin t_2} -$$
$$-\frac{2\eta}{N^2} \sum_{\vec{Q}} \sum_{\vec{\varkappa}} \sum_{\vec{k}} W(\vec{Q}) \varphi_n^*(\vec{\varkappa}) \varphi_n(\vec{k}) \exp(iz_1 \sin t_1 + iz_2 \sin t_2 + iz_3 \sin t_3), \; j = e, h. \tag{26}$$

In the same denotations the average e-h Hamiltonian is

$$\langle \psi_{bimex}(0, \eta, \varphi_n) | H_{e-h}^{LLL} | \psi_{bimex}(0, \eta, \varphi_n) \rangle = -\frac{4}{N} \sum_{\vec{Q}} \sum_{\vec{\varkappa}} W(\vec{Q}) \varphi_n^*(\vec{\varkappa}) \varphi_n(\vec{Q} - \vec{\varkappa}) -$$
$$-\frac{4}{N} \sum_{\vec{Q}} \sum_{\vec{\varkappa}} W(\vec{Q}) |\varphi_n(\vec{\varkappa})|^2 \cos(z_2 \sin t_2) + \tag{27}$$
$$+\frac{4\eta}{N^2} \sum_{\vec{Q}} \sum_{\vec{\varkappa}} \sum_{\vec{k}} W(\vec{Q}) \varphi_n^*(\vec{\varkappa}) \varphi_n(\vec{k}) \left[ \cos(z_2 \sin t_2) \cos(z_3 \sin t_3) + \cos(z_1 \sin t_1) \cos(z_3 \sin t_3) \right].$$

The average value of the full Coulomb interaction Hamiltonian (2) can be expressed as

$$\langle \psi_{bimex}(0, \eta, \varphi_n) | H_{Coul}^{LLL} | \psi_{bimex}(0, \eta, \varphi_n) \rangle = \frac{4}{N} \sum_{\vec{Q}} \sum_{\vec{\varkappa}} W(\vec{Q}) \varphi_n^*(\vec{\varkappa}) \times$$
$$\times \varphi_n(\vec{Q} - \vec{\varkappa})(e^{iz_2 \sin t_2} - 1) - \frac{4}{N} \sum_{\vec{Q}} \sum_{\vec{\varkappa}} W(\vec{Q}) |\varphi_n(\vec{\varkappa})|^2 \times$$
$$\times \cos(z_2 \sin t_2) + \frac{4\eta}{N^2} \sum_{\vec{Q}} \sum_{\vec{\varkappa}} \sum_{\vec{k}} W(\vec{Q}) \varphi_n^*(\vec{\varkappa}) \varphi_n(\vec{k}) \times \tag{28}$$
$$\times \left[ \cos(z_2 \sin t_2) \cos(z_3 \sin t_3) + \cos(z_1 \sin t_1) \cos(z_3 \sin t_3) - \right.$$
$$\left. - \exp(iz_1 \sin t_1 + iz_2 \sin t_2 + iz_3 \sin t_3) \right].$$

To obtain Eq. (28) we used well known formulas [20, 21]

$$e^{iz\sin t} = J_0(z) + 2\sum_{k=1}^{\infty} \left[ J_{2k}(z) \cos(2kt) + iJ_{2k-1}(z) \sin(2k-1)t \right];$$
$$\cos(z\sin t) = J_0(z) + 2\sum_{k=1}^{\infty} J_{2k}(z) \cos(2kt); \; e^{z\cos t} = I_0(z) + 2\sum_{k=1}^{\infty} I_k(z) \cos kt, \tag{29}$$

where $J_\upsilon(z)$ and $I_\upsilon(z)$ are the Bessel functions. Taking into account that the functions $W(\vec{Q})$, $\varphi_n(\vec{\varkappa})$ and $\varphi_n(\vec{k})$ depend only on the moduli $|\vec{Q}|$, $|\vec{\varkappa}|$ and $|\vec{k}|$, we obtain after integration over the angles $\varphi$, $\theta$ and $\psi$

$$\frac{1}{2\pi}\int_0^{2\pi}d\varphi\frac{1}{2\pi}\int_0^{2\pi}d\theta\frac{1}{2\pi}\int_0^{2\pi}d\psi e^{iz_1\sin t_1}e^{iz_2\sin t_2}e^{iz_3\sin t_3}=J_0(z_1)J_0(z_2)J_0(z_3)+2\sum_{k=1}^\infty J_{2k}(z_1)J_{2k}(z_2)J_{2k}(z_3);$$

$$\frac{1}{2\pi}\int_0^{2\pi}d\varphi\frac{1}{2\pi}\int_0^{2\pi}d\theta\frac{1}{2\pi}\int_0^{2\pi}d\psi\cos(z_i\sin t_i)\cos(z_3\sin t_3)=J_0(z_i)J_0(z_3),\ i=1,2;$$

$$\frac{1}{2\pi}\int_0^{2\pi}d\varphi\frac{1}{2\pi}\int_0^{2\pi}d\theta\frac{1}{2\pi}\int_0^{2\pi}d\psi\cos(z_i\sin t_i)=J_0(z_i),\ i=1,2,3;$$

$$\frac{1}{2\pi}\int_0^{2\pi}d\theta\frac{1}{2\pi}\int_0^{2\pi}d\psi e^{2\alpha xy\cos(\theta-\psi)}e^{ixy\sin(\theta-\psi)}=J_0(xy)I_0(2\alpha xy)+2\sum_{k=1}^\infty J_{2k}(xy)I_{2k}(2\alpha xy);$$

$$\frac{1}{2\pi}\int_0^{2\pi}d\theta\frac{1}{2\pi}\int_0^{2\pi}d\psi e^{2\alpha xy\cos(\theta-\psi)}=I_0(2\alpha xy). \qquad (30)$$

The angle integration excluding the trial function $\varphi_n(|\vec{x}-\vec{y}|)$ leads to the expression

$$\begin{aligned}
\langle\psi_{bimex}(0,\eta,\varphi_n)|H_{Coul}^{LLL}|\psi_{bimex}(0,\eta,\varphi_n)\rangle &= -4\left(\frac{e^2}{\varepsilon_0 l_0}\right)\int_0^\infty dy\, e^{-\frac{y^2}{2}}\times\\
&\times\int_0^\infty xdx|\varphi_n(x)|^2 J_0(xy)+4\left(\frac{e^2}{\varepsilon_0 l_0}\right)\int_0^\infty dy\, e^{-\frac{y^2}{2}}\int_0^\infty xdx\varphi_n^*(x)\frac{1}{2\pi}\int_0^{2\pi}d\theta\times\\
&\times\frac{1}{2\pi}\int_0^{2\pi}d\psi\varphi_n(|\vec{x}-\vec{y}|)\left(e^{ix\cdot y\sin(\theta-\psi)}-1\right)+4\eta\left(\frac{e^2}{\varepsilon_0 l_0}\right)\int_0^\infty dy\, e^{-\frac{y^2}{2}}\times\\
&\times\int_0^\infty xdx\varphi_n^*(x)\int_0^\infty zdz\varphi_n(z)\big(J_0(x\cdot y)J_0(x\cdot z)+J_0(x\cdot z)J_0(y\cdot z)-\\
&-J_0(x\cdot y)J_0(x\cdot z)J_0(y\cdot z)-2\sum_{k=1}^\infty J_{2k}(x\cdot y)J_{2k}(x\cdot z)J_{2k}(y\cdot z)\big),
\end{aligned} \qquad (31)$$

where $|\vec{x}-\vec{y}|=\sqrt{x^2+y^2-2xy\cos(\theta-\psi)}$.

The calculations of integrals in Eqs. (31) and (21) for the particular cases of variational wave functions $\varphi_2(x)$ and $\varphi_0(x)$ in the analytical form are presented in the Appendix.

## 4. The electron structure of the 2D quasi-bimagnetoexciton

As it was shown previously [10–12] the interaction of two 2D magnetoexcitons with wave vectors $\vec{k}=0$, composed of the electrons and holes lying on the LLLs vanishes because they look as two neutral compound particles. The interactions between them can appear under the influence of the ELLs as well as of the RSOC [13]. In the absence of these factors only two magnetoexcitons with the wave vectors $\vec{k} \neq 0$ can interact through the Coulomb forces and can form a quasi-molecular state. The molecule composed of two magnetoexcitons with antiparallel wave vectors $\vec{k}$ and $-\vec{k}$ has the structure of two antiparallel dipoles bound together and oriented with equal probability in any direction of the layer plane. Such possibility is achieved introducing the trial wave function of the relative motion of two magnetoexcitons in the frame of the molecule $\varphi_n(\vec{k})$, which depends on the modulus $k$.

Figs. 3 and 4 show the total energies of two bound 2D magnetoexcitons in units $2I_l$ for the variational wave functions $\varphi_2(x)$ and $\varphi_0(x)$, correspondingly. Such presentation facilitates the comparison of the obtained results with the energy of two free magnetoexcitons with wave vectors $\vec{k} = 0$.

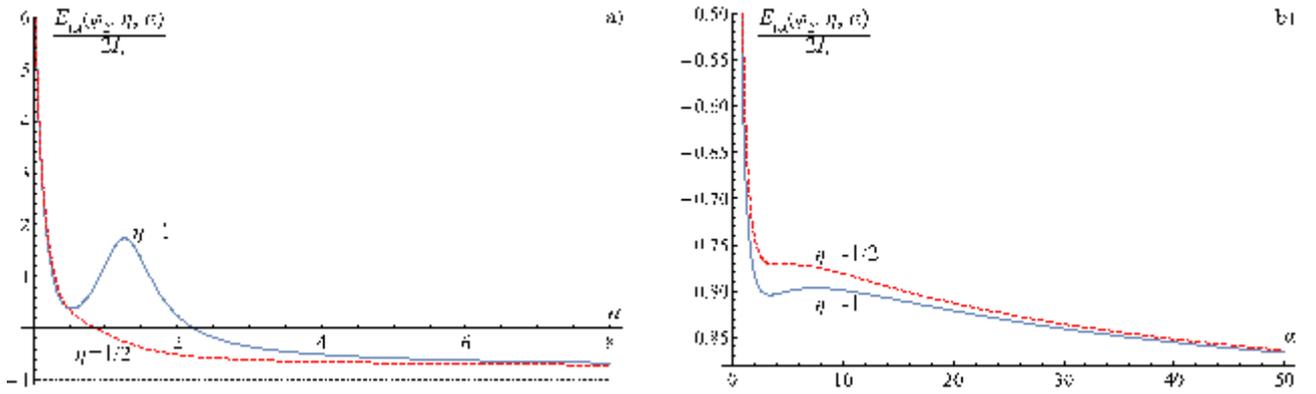

**Fig. 3.** Total energies of the bound states of two 2D magnetoexcitons with wave vectors $\vec{k}$ and $-\vec{k}$, with different spin structures $\eta = \pm 1, \pm 1/2$ and with the variational wave function $\varphi_2(k)$, in dependence on the parameter $\alpha$: a) $\eta = 1, 1/2$; b) $\eta = -1, -1/2$. The total energies are normalized to the value $2I_l$, where $I_l$ is the ionization potential of a free magnetoexciton with wave vector $\vec{k} = 0$. The energy of two free magnetoexcitons with $k = 0$ is represented by the (dotted) line.

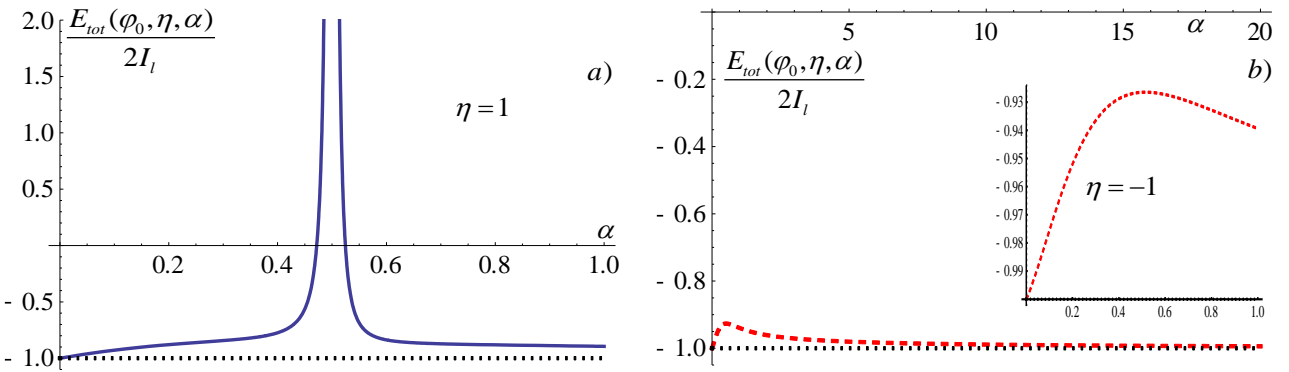

**Fig. 4.** The total energy of two bound 2D magnetoexcitons with the trial wave function $\varphi_0(x)$ for different spin structures $\eta = \pm 1$ in dependence on the parameter $\alpha$: a) $\eta = 1$, b) $\eta = -1$.

The numerical calculations made for the function $\varphi_2(\vec{k}) = (8\alpha^3)^{1/2}(kl_0)^2 e^{-\alpha(kl_0)^2}$ allow to obtain the full energies of the bound states in dependence on the parameter $\alpha$ of the trial wave function in four cases with $\eta = \pm 1, \pm 1/2$ corresponding to two electrons and holes spin structures. In four spin configurations the full energies of the bound states are greater than $-2I_l$ for all values of $\alpha$. All these states are unstable as regards the dissociation in two free magnetoexcitons with $\vec{k}=0$. In spite of this, a deep metastable bound state with the activation barrier comparable with two magnetoexciton ionization potentials $2I_l$ in the case $\eta=1$ and $\alpha=0.5$ is revealed. In the opposite case $\eta=-1$ and $\alpha=3.4$ only a shallow metastable bound state is observed. As one can see from Fig. 4, all bound states obtained with the trial functions $\varphi_0(x)$ are unstable.

Recall that in for the hydrogen molecule [23] the two-electron wave function written as a product of the orbital wave function depending on the electrons space coordinates and the spinor-type wave function depending on their spin coordinates can be either combination of the symmetric orbital wave function multiplied by the antisymmetric spinor function gives corresponding to the singlet $^1\Sigma_g^+$ strongly bound molecular state, which is the ground state of H$_2$ molecule. In the antisymmetric $^3\Sigma_u^+$ case the energy between atoms decreases monotonically as the distance between the nuclei increases, corresponding to the mutual repulsion of the two atoms. The situation becomes completely different in the case of hydrogen atoms in a strong magnetic field, when the triplet term becomes a ground state with possibility of a deep potential well in interatomic interaction [3, 4]. Comparing the obtained results for four spin configurations we can notice that neither ortho-ortho configuration or para-para configuration proved to be less favorable for the formation of the metastable bound state.

The metastable bound state with a lifetime of about few picoseconds, and possibility of a new luminescence band due to the radiative recombination of one electron-hole pair and the conversion of the metastable bound state into the para-magnetoexciton and the emission of the photon was predicted in [14, 15]. The new luminescence band should be at the higher energy side compared to the para-magnetoexciton luminescence line.

## 5. Discussion and Conclusions

Bound states of two magnetoexcitons with opposite wave vectors $\vec{k}$ and $-\vec{k}$ were investigated in the lowest Landau levels approximation, neglecting by the influence of the excited Landau levels. The electrons and the heavy holes are situated on the LLLs with the cyclotron energies greater than the binding energy of the 2D Wannier-Mott exciton. The spin states of two electrons and the effective spins of two heavy holes were combined to form states characterized by the parameter $\eta$ with four different values $\eta=\pm 1, \pm 1/2$ corresponding to four different spin structures: triplet-triplet, singlet-singlet, ortho-ortho and para-para. Each magnetoexciton with wave vector $\vec{k} \neq 0$ looks as an electric dipole with the length of the arm between the electron and the hole equal to $d = k l_0^2$. The arm is oriented in-plane perpendicular to the wave vector $\vec{k}$. The bound state is formed by two quickly changing dipoles with antiparallel arms and with the total wave vector equal to zero. The bound pair of two dipoles is oriented arbitrary in the plane of the layer and it is characterized by the trial wave function of the relative motion $\varphi_n(\vec{k})$, which depends only on the modulus $|\vec{k}|$. The numerical calculations used the trial wave function $\varphi_2(k) = (8\alpha^3)^{1/2}(k l_0)^2 e^{-\alpha(k l_0)^2}$ and have shown the absence of the stable molecular bound states in all four spin orientations $\eta = \pm 1, \pm 1/2$. Only metastable bound states were revealed. One of them is a deep bound state for $\eta = 1$ and $\alpha = 0.5$ characterized by the activation barrier comparable with two magnetoexciton ionization potentials $2I_l$. For $\eta = -1$ and $\alpha = 3.4$ only a shallow metastable bound state can be formed. The papa-para and ortho-ortho magnetexcitons do not have considerable barriers and cannot manifest similar metastable bound states, as in the case of triplet-triplet spin combination. In the case of the magnetoexcitons the bound quasi-molecular states are formed by the four components, namely by two electrons and by two holes. As in the case of the hydrogen molecule the Coulomb interaction energies of the singlet-singlet spin combination corresponding to $\eta = -1$ are situated on the lower energy branch than in the case of the triplet-triplet spin combination corresponding to $\eta = 1$. But in the LLLs approximation all these states are unstable. As it was mentioned above, one deep metastable bound state was revealed in the case of the triplet-triplet spin structure with the trial wave function $\varphi_2(x)$ and $\alpha = 0.5$. In this case we can suppose the existence of the quasi-stable bound state and the formation of the quasi-bimagnetoexciton.

## Acknowledgements

IVP thanks S. Moskalenko for useful discussions. This work was performed in the frame of the institutional project #15.817.02.05F of the Institute of Applied Physics, Republic of Moldova.

## Appendix

To calculate integrals (31) we consider first the trial wave function $\varphi_2(x)$ in the form

$$\varphi_2\left(|\vec{x} - \vec{y}|\right) = \left(8\alpha^3\right)^{1/2} |\vec{x} - \vec{y}|^2 e^{-\alpha|\vec{x}-\vec{y}|^2} = \left(8\alpha^3\right)^{1/2} e^{-\alpha(x^2+y^2)} \times$$
$$\times \left[ x^2 + y^2 - 2xy \frac{\partial}{\partial(2\alpha xy)} \right] \left[ I_0(2\alpha xy) + 2\sum_{k=1}^{\infty} I_k(2\alpha xy) \cos(k(\theta - \varphi)) \right]. \quad (A1)$$

Using the integrals (30), one can obtain the expression

$$\frac{1}{2\pi}\int_0^{2\pi} d\theta \frac{1}{2\pi}\int_0^{2\pi} d\psi \varphi_2(|\vec{x}-\vec{y}|)\left(e^{ixy\sin(\theta-\psi)}-1\right) = (8\alpha^3)^{1/2} e^{-\alpha(x^2+y^2)} \times$$

$$\times \left\{ (x^2+y^2)\left[ (J_0(xy)-1)I_0(2\alpha xy) + \sum_{k=1}^{\infty} J_{2k}(xy)I_{2k}(2\alpha xy) \right] - \right.$$

$$-2xy\left[ (J_0(xy)-1)I_1(2\alpha xy) + \sum_{k=1}^{\infty} J_{2k}(xy)I_{2k+1}(2\alpha xy) \right] -$$

$$\left. -\frac{2}{\alpha}\sum_{k=1}^{\infty} k \cdot J_{2k}(xy)I_{2k}(2\alpha xy) \right\}.$$
(A2)

Here the derivatives of the Bessel functions $I_n(z)$ and $J_n(z)$ of the integer order were used [20–22]

$$\frac{dI_n(z)}{dz} = \frac{n}{z}I_n(z) + I_{n+1}(z); \quad \frac{dJ_n(z)}{dz} = \frac{n}{z}J_n(z) - J_{n+1}(z).$$
(A3)

Taking into account Eq. (A2), the second term in the right hand side of the average value (31) can be transcribed in the following way

$$\varepsilon_2(\varphi_2,\eta,\alpha) = 4\left(\frac{e^2}{\varepsilon_0 l_0}\right)\int_0^{\infty} dy\, e^{-\frac{y^2}{2}} \int_0^{\infty} xdx \varphi_2^*(x) \frac{1}{2\pi}\int_0^{2\pi} d\theta \frac{1}{2\pi}\int_0^{2\pi} d\psi \times$$

$$\times \varphi_2(|\vec{x}-\vec{y}|)\left(e^{ixy\sin(\theta-\psi)}-1\right) = 4\left(\frac{e^2}{\varepsilon_0 l_0}\right)(8\alpha^3) \times$$

$$\times \left\{ \int_0^{\infty} dy\, e^{-y^2\left(\frac{1}{2}+\alpha\right)} \int_0^{\infty} dx\, x^5 e^{-2\alpha x^2} J_0(x \cdot y) I_0(2\alpha xy) + \int_0^{\infty} dy\, y^2 e^{-y^2\left(\frac{1}{2}+\alpha\right)} \int_0^{\infty} dx\, x^3 e^{-2\alpha x^2} J_0(xy) I_0(2\alpha xy) - \right.$$

$$-\int_0^{\infty} dy\, e^{-y^2\left(\frac{1}{2}+\alpha\right)} \int_0^{\infty} dx\, x^5 e^{-2\alpha x^2} I_0(2\alpha xy) - 2\int_0^{\infty} dy \cdot y\, e^{-y^2\left(\frac{1}{2}+\alpha\right)} \int_0^{\infty} dx\, x^4 e^{-2\alpha x^2} J_0(xy) I_1(2\alpha xy) -$$

$$-\int_0^{\infty} dy\, y^2 e^{-y^2\left(\frac{1}{2}+\alpha\right)} \int_0^{\infty} dx\, x^3 e^{-2\alpha x^2} I_0(2\alpha xy) + 2\int_0^{\infty} dy \cdot y\, e^{-y^2\left(\frac{1}{2}+\alpha\right)} \int_0^{\infty} dx\, x^4 e^{-2\alpha x^2} I_1(2\alpha xy) +$$

$$+2\int_0^{\infty} dy\, e^{-y^2\left(\frac{1}{2}+\alpha\right)} \int_0^{\infty} dx \cdot x^5 e^{-2\alpha x^2} \sum_{k=1}^{\infty} J_{2k}(xy) I_{2k}(2\alpha xy) -$$

$$-4\int_0^{\infty} dy \cdot y\, e^{-y^2\left(\frac{1}{2}+\alpha\right)} \int_0^{\infty} dx\, x^4 e^{-2\alpha x^2} \sum_{k=1}^{\infty} J_{2k}(xy) I_{2k+1}(2\alpha xy) +$$

$$+2\int_0^{\infty} dy\, y^2 e^{-y^2\left(\frac{1}{2}+\alpha\right)} \int_0^{\infty} dx\, x^3 e^{-2\alpha x^2} \sum_{k=1}^{\infty} J_{2k}(xy) I_{2k}(2\alpha xy) -$$

$$\left. -\frac{4}{\alpha}\int_0^{\infty} dy\, e^{-y^2\left(\frac{1}{2}+\alpha\right)} \int_0^{\infty} dx\, x^3 e^{-2\alpha x^2} \sum_{k=1}^{\infty} k J_{2k}(xy) I_{2k}(2\alpha xy) \right\}.$$
(A4)

Similar calculations using the trial wave function $\varphi_0(k) = \sqrt{4\alpha}\, e^{-\alpha k^2 l_0^2}$ result in

$$\varepsilon_2(\varphi_0,\eta,\alpha) = -16\alpha\left(\frac{e^2}{\varepsilon_0 l_0}\right)\int_0^\infty dy e^{-y^2\left(\frac{1}{2}+\alpha\right)}\int_0^\infty dxxe^{-2\alpha x^2} \times$$

$$\times I_0(2\alpha xy) + 16\alpha\left(\frac{e^2}{\varepsilon_0 l_0}\right)\int_0^\infty dy e^{-y^2\left(\frac{1}{2}+\alpha\right)}\int_0^\infty dxxe^{-2\alpha x^2} \times$$

$$\times\left[J_0(xy)I_0(2\alpha xy) + \sum_{n=1}^\infty J_{2n}(xy)I_{2n}(2\alpha xy)\right] = \quad (A5)$$

$$= -4I_l \frac{1}{\sqrt{1+\alpha}} + 2I_l\sqrt{\frac{2}{\pi}}I_0^{1/2}(q',c') + 4I_l\sqrt{\frac{2}{\pi}}\sum_{n=1}^\infty I_{2n}^{1/2}(q',c');$$

$$q' = \frac{4\alpha^2 + 4\alpha + 1}{8a}; \quad c' = \frac{1}{2}.$$

The third contribution to the average value (31) is

$$\varepsilon_3(\varphi_2,\eta,\alpha) = 4\eta\left(\frac{e^2}{\varepsilon_0 l_0}\right)\int_0^\infty dy e^{-\frac{y^2}{2}}\int_0^\infty xdx\varphi_2^*(x)\int_0^\infty zdz\varphi_2(z)\big(J_0(x\cdot y)J_0(x\cdot z) + J_0(x\cdot z)J_0(y\cdot z) -$$

$$- J_0(x\cdot y)J_0(x\cdot z)J_0(y\cdot z) - 2\sum_{k=1}^\infty J_{2k}(x\cdot y)J_{2k}(x\cdot z)J_{2k}(y\cdot z)\big) = \quad (A6)$$

$$= 4\eta\left(\frac{e^2}{\varepsilon_0 l_0}\right)(8\alpha^3)\int_0^\infty dy e^{-\frac{y^2}{2}}\int_0^\infty dxx^3 e^{-\alpha x^2}\int_0^\infty dzz^3 e^{-\alpha z^2}\big[J_0(x\cdot y)J_0(x\cdot z) + J_0(x\cdot z)J_0(y\cdot z) -$$

$$- J_0(x\cdot y)J_0(x\cdot z)J_0(y\cdot z) - 2\sum_{k=1}^\infty J_{2k}(x\cdot y)J_{2k}(x\cdot z)J_{2k}(y\cdot z)\big].$$

The first term in the average value (31), as well as the overlapping integrals $L_n(\alpha)$ can be calculated analytically

$$\varepsilon_1(\varphi_2,\eta,\alpha) = -4\left(\frac{e^2}{\varepsilon_0 l_0}\right)\int_0^\infty dy e^{-\frac{y^2}{2}}\int_0^\infty dxx|\varphi_2(x)|^2 J_0(xy) = -4I_l\cdot\frac{\sqrt{4\alpha}}{\sqrt{1+4\alpha}}\left[1 - \frac{1}{1+4\alpha} + \frac{3}{8}\frac{1}{(1+4\alpha)^2}\right];$$

(A7)

$$I_l = \left(\frac{e^2}{\varepsilon_0 l_0}\right)\sqrt{\frac{\pi}{2}}; \quad L_2(\alpha) = \frac{2\alpha^2 - \frac{1}{2}}{\left(\alpha + \frac{1}{4\alpha}\right)^3}.$$

Here $I_l$ is the ionization potential of the 2D magnetoexciton with wave vector $k_\| = 0$. It is convenient to rewrite Eq. (A7) in the form

$$\frac{E_1(\varphi_2,\eta,\alpha)}{2I_l} = \frac{\varepsilon_1(\varphi_2,\eta,\alpha)}{4I_l(1-\eta L_2(\alpha))} = -\frac{\sqrt{4\alpha}\left[1 - \frac{1}{1+4\alpha} + \frac{3}{8}\frac{1}{(1+4\alpha)^2}\right]}{\sqrt{1+4\alpha}\left[1 - \frac{\eta(2\alpha^2 - 1/2)}{(\alpha + 1/(4\alpha))^3}\right]}. \quad (A8)$$

For the trial function $\varphi_0(x)$ we obtain more simple expression instead of Eq. (A7)

$$\frac{E_1(\varphi_0,\eta,\alpha)}{2I_l} = \frac{\varepsilon_1(\varphi_0,\eta,\alpha)}{4I_l(1-\eta L_0(\alpha))} = -\frac{\sqrt{4\alpha}}{\sqrt{1+4\alpha}}\frac{1}{(1-\eta L_0(\alpha))}. \tag{A9}$$

In Eq. (A4) for $\varepsilon_2(\varphi_2,\eta,\alpha)$ there are three integrals containing only one modified Bessel function $I_\nu(2\alpha xy)$ with $\nu = 0,1$, which depend on the parameter $\alpha$ of the variational wave function $\varphi_2(x)$

$$\begin{aligned}
I_1 &= -\int_0^\infty dy\, e^{-y^2\left(\frac{1}{2}+\alpha\right)}\int_0^\infty dx\, x^5 e^{-2\alpha x^2} I_0(2\alpha xy) = -\frac{\sqrt{2\pi}}{128\alpha^3}\cdot\frac{(8+24\alpha+19\alpha^2)}{(1+\alpha)^{5/2}}, \\
I_2 &= -\int_0^\infty dy\, y^2 e^{-y^2\left(\frac{1}{2}+\alpha\right)}\int_0^\infty dx\, x^3 e^{-2\alpha x^2} I_0(2\alpha xy) = -\frac{\sqrt{2\pi}}{32\alpha^2}\cdot\frac{(2+5\alpha)}{(1+\alpha)^{5/2}}, \\
I_3 &= 2\int_0^\infty dy\cdot y\cdot e^{-y^2\left(\frac{1}{2}+\alpha\right)} I_1(2\alpha xy) = \frac{\sqrt{2\pi}}{32\alpha^2}\cdot\frac{(4+7\alpha)}{(1+\alpha)^{5/2}}
\end{aligned} \tag{A10}$$

Another three integrals $I_4$, $I_5$ and $I_6$ containing the products of two Bessel functions of the type $J_0(bx)I_0(cx)$ and $J_0(bx)I_1(cx)$ were also calculated analytically:

$$\begin{aligned}
I_4 &= \int_0^\infty dy\, e^{-y^2\left(\frac{1}{2}+\alpha\right)}\int_0^\infty dx\, x^5 e^{-2\alpha x^2} J_0(xy)I_0(2\alpha xy) = \\
&= \frac{1}{32\alpha^4}\left\{3I_0^{1/2}(q,c) + \frac{3}{8\alpha}(4\alpha^2-1)I_0^{3/2}(q,c) + \right.\\
&\quad \left. + \frac{(4\alpha^2-1)}{128\alpha^2}I_0^{5/2}(q,c) - \frac{7}{4}I_1^{3/2}(q,c) - \frac{(4\alpha^2-1)}{16\alpha}I_1^{5/2}(q,c) + \frac{1}{8}I_2^{5/2}(q,c)\right\},
\end{aligned}$$

$$\begin{aligned}
I_5 &= \int_0^\infty dy\, y^2 e^{-y^2\left(\frac{1}{2}+\alpha\right)}\int_0^\infty dx\, x^3 e^{-2\alpha x^2} J_0(xy)I_0(2\alpha xy) = \\
&= \frac{1}{16\alpha^2}\left\{I_0^{3/2}(q,c) + \frac{(4\alpha^2-1)}{8\alpha}I_0^{5/2}(q,c) - \frac{1}{2}I_1^{5/2}(q,c)\right\},
\end{aligned}$$

$$\begin{aligned}
I_6 &= -2\int_0^\infty dy\, y\, e^{-y^2\left(\frac{1}{2}+\alpha\right)}\int_0^\infty dx\, x^4 e^{-2\alpha x^2} J_0(xy)I_1(2\alpha xy) = \\
&= -\frac{1}{16\alpha^3}\left\{I_0^1(q,c) + \frac{(4\alpha^2-1)}{16\alpha}I_0^2(q,c) - \frac{1}{4}I_1^2(q,c) - \right.\\
&\quad \left. -\frac{3}{2}I_1^{3/2}(q,c) - \frac{(4\alpha^2-1)}{16\alpha}I_1^{5/2}(q,c) + \frac{1}{4}I_2^{5/2}(q,c)\right\}.
\end{aligned} \tag{A11}$$

Here were used notations [22]

$$I_0^1(q,c) = \frac{1}{(q^2+c^2)^{1/2}};\ I_0^2(q,c) = \frac{q}{(q^2+c^2)^{3/2}};\ I_1^2(q,c) = \frac{c}{(q^2+c^2)^{3/2}};\ I_0^{1/2}(q,c) = 2\sqrt{\frac{(1-2k^2)}{\pi q}}K(k),$$

$$I_0^{3/2}(q,c) = \sqrt{\frac{(1-2k^2)^3}{\pi q^3}}\left[2E(k) - K(k)\right],$$

$$I_0^{5/2}(q,c) = \frac{1}{2}\sqrt{\frac{(1-2k^2)^5}{\pi q^5}}\left[8(1-2k^2)E(k)-(5-8k^2)K(k)\right],$$

$$I_1^{3/2}(q,c) = \frac{1}{k}\sqrt{\frac{(1-2k^2)^3}{\pi q^3(1-k^2)}}\left[(1-k^2)K(k)-(1-2k^2)E(k)\right],$$

$$I_1^{5/2}(q,c) = \frac{1}{2k}\sqrt{\frac{(1-2k^2)^5}{\pi q^5(1-k^2)}}\left[(1-k^2)(1-8k^2)K(k)-(1-16k^2+16k^4)E(k)\right],$$

$$I_2^{5/2}(q,c) = \frac{1}{2k^2(1-k^2)}\sqrt{\frac{(1-2k^2)^5}{\pi q^5}}\left[(2+5k^2-8k^4)\times\right.$$

$$\left.\times(1-k^2)K(k)-2(1-2k^2)(1+4k^2-4k^4)E(k)\right].$$

(A12)

The complete elliptic integrals of the first and second kinds $K(k)$ and $E(k)$ depend on the modulus $k$, which in its turn depends on the parameters $q$ and $c$ in the way

$$k = \frac{1}{\sqrt{2}}\left(1-\frac{q}{\sqrt{q^2+c^2}}\right)^{1/2}, \quad q = \frac{1+4\alpha+4\alpha^2}{8\alpha}; \quad c = \frac{1}{2}. \tag{A13}$$

In the range of the small values $k \ll 1$, the series expansions of the functions $K(k)$ and $E(k)$ can be used

$$K(k)\underset{k\to 0}{\approx}\frac{\pi}{2}\left(1+\frac{k^2}{4}+\frac{9}{64}k^4\right); \quad E(k)\underset{k\to 0}{\approx}\frac{\pi}{2}\left(1-\frac{k^2}{4}-\frac{3}{64}k^4\right); \tag{A14}$$

In this limit the apparent singular expressions in formulas (A10) can be transformed in the regular forms

$$\frac{1}{k}\left[(1-k^2)K(k)-(1-2k^2)E(k)\right] \approx \frac{3}{4}\pi k\left(1-\frac{3}{8}k^2\right);$$

$$\frac{1}{k}\left[(1-k^2)(1-8k^2)K(k)-(1-16k^2+16k^4)E(k)\right] \approx \frac{15}{4}\pi k\left(1-\frac{15}{8}k^2\right); \tag{A15}$$

$$\frac{1}{k^2}\left[(2+5k^2-8k^4)(1-k^2)K(k)-2(1-2k^2)(1+4k^2-4k^4)E(k)\right] \approx \frac{105}{16}\pi k^2; \quad k \to 0$$

Using these expressions we obtain simplified formulas in the limit $k \to 0$

$$I_1^{3/2}(q,c) = \frac{3}{4}\pi k\sqrt{\frac{(1-2k^2)^3}{\pi q^3(1-k^2)}}\left(1-\frac{3}{8}k^2\right); I_1^{5/2}(q,c) = \frac{15}{8}\pi k\sqrt{\frac{(1-2k^2)^5}{\pi q^5(1-k^2)}}\left(1-\frac{15}{8}k^2\right)$$

$$I_2^{5/2}(q,c) = \frac{105}{32}\pi\frac{k^2}{(1-k^2)}\sqrt{\frac{(1-2k^2)^5}{\pi q^5}}; k \to 0. \tag{A16}$$

The value of integrals $I_\nu^\beta$ for different values of the parameter $\alpha$ for $q=(1+4\alpha+4\alpha^2)/(8\alpha)$ and $c=1/2$ are represented in the figure 5.

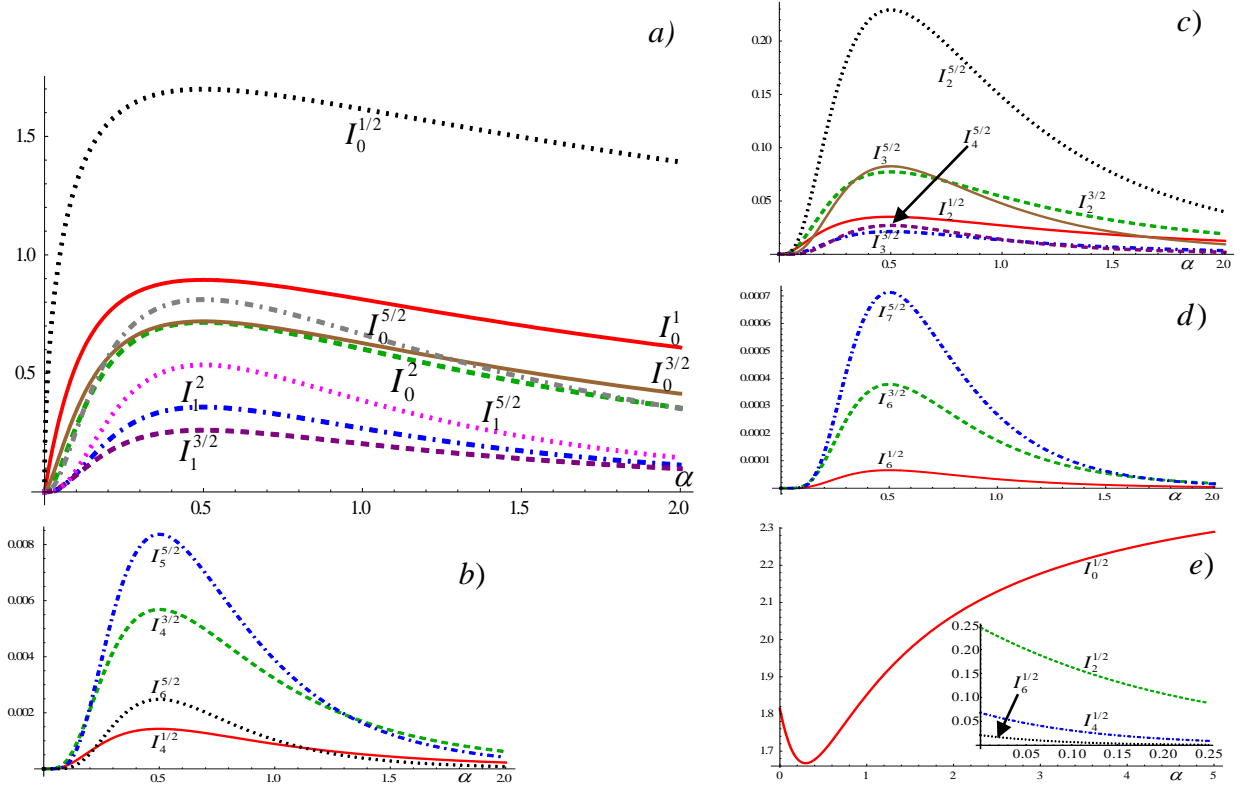

**Fig. 5.** The table integrals $I_\nu^\beta$ as function of $\alpha$ for different indices $\nu$ and $\beta$. a), b), c), d): $q=(4\alpha^2+4\alpha+1)/(8\alpha)$, $c=1/2$; e): $q=(1+4\alpha+4\alpha^2)/(2(1+4\alpha^2))$, $c=1/(1+4\alpha^2)$.

The third contribution $\varepsilon_3(\varphi_2,\eta,\alpha)$ in Eq. (A6) is determined by integrals $I_7, I_8$ and $I_9$. The integral $I_7$ equals to

$$I_7 = \int_0^\infty dy\, e^{-\frac{y^2}{2}} \int_0^\infty dx\, x^3 e^{-\alpha x^2} \int_0^\infty dz\, z^3 e^{-\alpha z^2} \left(J_0(xy)J_0(xz)+J_0(xz)J_0(yz)\right) = $$
$$= \frac{4\sqrt{\pi}(4\alpha^2-1)}{(4\alpha^2+1)^3 q^{1/2}} - \frac{2\alpha\sqrt{\pi}(4\alpha^2-3)}{(4\alpha^2+1)^4 q^{3/2}} - \frac{3\alpha^2\sqrt{\pi}}{4(4\alpha^2+1)^4 q^{5/2}}. \quad (A17)$$

It was calculated taking into account that the product of two Bessel functions $J_0(xz)J_0(yz)$ can be transformed into the expression $J_0(xy)J_0(xz)$ by the interchange of the variables $x \rightleftarrows z$, and using the integral [22]:

$$\int_0^\infty dz\, z\, e^{-pz^2} J_0(xz) = \frac{1}{2p} e^{-\frac{x^2}{4p}}. \quad (A18)$$

Two integrals $I_8$ and $I_9$ contain three Bessel functions

$$I_8 = -\int_0^\infty dy\, e^{-\frac{y^2}{2}} \int_0^\infty dx\, x^3 e^{-\alpha x^2} \int_0^\infty dz\, z^3 e^{-\alpha z^2} J_0(xy) J_0(xz) J_0(yz), \quad (A19)$$

and

$$I_9 = -2\sum_{n=1}^{\infty}\int_0^{\infty} dy\, e^{-\frac{y^2}{2}} \int_0^{\infty} dx\, x^3 e^{-\alpha x^2} \int_0^{\infty} dz\, z^3 e^{-\alpha z^2} \cdot J_{2n}(xy) J_{2n}(xz) J_{2n}(yz). \quad (A20)$$

They are calculated using the formulas [20–22]

$$\int_0^{\infty} dz\, z e^{-\alpha z^2} J_0(xz) J_0(yz) = \frac{1}{2\alpha} e^{-\frac{(x^2+y^2)}{4\alpha}} I_0\left(\frac{xy}{2\alpha}\right),$$

$$\int_0^{\infty} dx\, x e^{-px^2} J_\upsilon(bx) I_\upsilon(cx) = \frac{1}{2p} \exp\left(\frac{c^2 - b^2}{4p}\right) J_\upsilon\left(\frac{bc}{2p}\right), \quad (A21)$$

$$\int_0^{\infty} x^{\beta-1} e^{-qx} J_\upsilon(cx) = I_\upsilon^\beta(q,c).$$

The third contribution $\varepsilon_3(\varphi_n, \eta, \alpha)$, described by the formula (A6), in the case of the wave function $\varphi_0(x)$ looks as

$$\varepsilon_3(\varphi_0, \eta, \alpha) = 16\eta\alpha\left(\frac{e^2}{\varepsilon_0 l_0}\right)\left[I_7^0 + I_8^0 + I_9^0\right]. \quad (A22)$$

These integrals are more simple than in the case of the trial wave function $\varphi_2(x)$ and equal to

$$I_7^0 = 2\int_0^{\infty} dy\, e^{-\frac{y^2}{2}} \int_0^{\infty} dx\, x e^{-\alpha x^2} \int_0^{\infty} dz\, z e^{-\alpha z^2} J_0(xy) J_0(xz) = \frac{\sqrt{2\pi}}{2(1+4\alpha^2)} I_0^{1/2}(q,c),$$

$$I_8^0 = -\int_0^{\infty} dy\, e^{-\frac{y^2}{2}} \int_0^{\infty} dx\, x e^{-\alpha x^2} \int_0^{\infty} dz\, z e^{-\alpha z^2} J_0(xy) J_0(xz) J_0(yz) = -\frac{1}{2}\frac{1}{(1+4\alpha^2)} I_0^{1/2}(q,c),$$

$$I_9^0 = -2\sum_{n=1}^{\infty}\int_0^{\infty} dy\, e^{-\frac{y^2}{2}} \int_0^{\infty} dx\, x e^{-\alpha x^2} \int_0^{\infty} dz\, z e^{-\alpha z^2} J_{2n}(xy) J_{2n}(xz) J_{2n}(yz) = -\frac{1}{(1+4\alpha^2)}\sum_{n=1}^{\infty} I_{2n}^{1/2}(q,c), \quad (A23)$$

$$q = \frac{1+4\alpha+4\alpha^2}{2(1+4\alpha^2)}; c = \frac{1}{(1+4\alpha^2)}.$$

There are still four double integrals $I_{10} - I_{13}$ in the composition of the expression (A4). They were calculated analytically exactly below. In all of them as the first step was used the table integral [22]

$$Z_{2n}(p,b,c) = \int_0^{\infty} dx\, x e^{-px^2} J_{2n}(bx) I_{2n}(cx) = \frac{1}{2p} e^{\frac{(c^2-b^2)}{4p}} J_{2n}\left(\frac{bc}{2p}\right) \quad (A24)$$

and its derivatives

$$-\frac{d}{dp}Z_{2n}(p,b,c) = \frac{1}{2p^2}\exp\left(\frac{c^2-b^2}{4p}\right)\left[\left(2n+1+\frac{c^2-b^2}{4p}\right)J_{2n}\left(\frac{bc}{2p}\right)-\right.$$

$$\left.-\frac{bc}{2p}J_{2n+1}\left(\frac{bc}{2p}\right)\right] = \int_0^\infty dx\, x^3 e^{-px^2} J_{2n}(bx)I_{2n}(cx);$$

$$\frac{d^2}{dp^2}Z_{2n}(p,b,c) = \int_0^\infty dx\, x^5 e^{-px^2} J_{2n}(bx)I_{2n}(c,x) =$$

$$= \frac{1}{p^3}e^{\frac{c^2-b^2}{4p}}\left\{J_{2n}\left(\frac{bc}{2p}\right)\left[\left(1+n+\frac{c^2-b^2}{8p}\right)\left(1+2n+\frac{c^2-b^2}{4p}\right)+\frac{c^2-b^2}{8p}\right]-\right.$$

$$\left.-\frac{bc}{4p}\left(4n+5+\frac{c^2-b^2}{2p}\right)J_{2n+1}\left(\frac{bc}{2p}\right)+\frac{b^2c^2}{8p}J_{2n+2}\left(\frac{bc}{2p}\right)\right\};$$

$$\left[-\frac{\partial^2}{\partial p\partial c}+\frac{2n}{c}\frac{\partial}{\partial p}\right]Z_{2n}(p,b,c) = \int_0^\infty dx\, x^4 e^{-px^2} J_{2n}(bx)I_{2n+1}(cx) =$$

$$= \frac{1}{2p^2}e^{\frac{c^2-b^2}{4p}}\left\{\frac{c}{2p}\left[2(n+1)+\frac{c^2-b^2}{4p}\right]J_{2n}\left(\frac{bc}{2p}\right)-\right.$$

$$\left.-\left[\frac{bc^2}{4p^2}+\frac{b}{2p}\left(2n+3+\frac{c^2-b^2}{4p}\right)\right]J_{2n+1}\left(\frac{bc}{2p}\right)+\frac{b^2c}{4p^2}J_{2n+2}\left(\frac{bc}{2p}\right)\right\}. \quad (A25)$$

In all these formulas it is necessary to substitute the parameters $p=2x$, $b=y$ and $c=2\alpha y$. They lead to the expressions $(c^2-b^2)/(4p) = y^2(4\alpha^2-1)/(8\alpha)$, $(bc)/(2p) = y^2/2$; $c/(2p) = y/2$; $(bc^2)/(4p^2) = y^3/4$, $(b^2c)/(4p^2) = y^3/(8\alpha)$, $1/(2p^2) = 1/(8\alpha^2)$. Formulas (A25) make possible to calculate the integral

$$\int_0^\infty dx\, x^4 e^{-2\alpha x^2} J_{2n}(xy)I_{2n+1}(2\alpha xy) = \frac{1}{8\alpha^2}e^{\frac{(4\alpha^2-1)}{8\alpha}y^2} \times$$

$$\times\left\{\frac{y}{2}\left[2(n+1)+\frac{(4\alpha^2-1)}{8\alpha}y^2\right]J_{2n}\left(\frac{y^2}{2}\right)-\left[\frac{y}{4\alpha}\left(2n+3+\frac{(4\alpha^2-1)}{8\alpha}y^2\right)\right]\times\right. \quad (A26)$$

$$\left.\times J_{2n+1}\left(\frac{y^2}{2}\right)+\frac{y^3}{8\alpha^2}J_{2n+2}\left(\frac{y^2}{2}\right)\right\}.$$

The following integration over the variable $y$, using the table integral [22]

$$\int_0^\infty z^{\alpha-1} e^{-qz} J_\nu(cz)dz = I_\nu^\alpha(q,c), \quad (A27)$$

gives the possibility to calculate the last four double integrals as follows:

$$I_{10} = 2\sum_{n=1}^{\infty} \int_0^{\infty} dy y^2 e^{-y^2\left(\frac{1}{2}+\alpha\right)} \int_0^{\infty} dx x^3 e^{-2\alpha x^2} J_{2n}(xy) I_{2n}(2\alpha xy) =$$

$$= \frac{1}{8\alpha^2} \sum_{n=1}^{\infty} \left[ (2n+1) I_{2n}^{3/2}(q,c) + \frac{(4\alpha^2-1)}{8\alpha} I_{2n}^{5/2}(q,c) - \frac{1}{2} I_{2n+1}^{5/2}(q,c) \right],$$

$$I_{11} = -\frac{4}{\alpha} \sum_{n=1}^{\infty} n \int_0^{\infty} dy e^{-y^2\left(\frac{1}{2}+\alpha\right)} \int_0^{\infty} dx x^3 e^{-2\alpha x^2} J_{2n}(xy) I_{2n}(2\alpha xy) =$$

$$-\frac{1}{4\alpha^3} \sum_{n=1}^{\infty} n \left[ (1+2n) I_{2n}^{1/2}(q,c) + \frac{(4\alpha^2-1)}{8\alpha} I_{2n}^{3/2}(q,c) - \frac{1}{2} I_{2n+1}^{3/2}(q,c) \right];$$

$$I_{12} = 2\sum_{n=1}^{\infty} \int_0^{\infty} dy e^{-y^2\left(\frac{1}{2}+\alpha\right)} \int_0^{\infty} dx x^5 e^{-2\alpha x^2} J_{2n}(xy) I_{2n}(2\alpha xy) =$$

$$= \frac{1}{8\alpha^3} \sum_{n=1}^{\infty} \left[ (1+n)(1+2n) I_{2n}^{1/2}(q,c) + (1+n) \frac{(4\alpha^2-1)}{4\alpha} I_{2n}^{3/2}(q,c) + \right.$$

$$\left. + \frac{(4\alpha^2-1)^2}{128\alpha^2} I_{2n}^{5/2}(q,c) - \frac{(4n+5)}{4} I_{2n+1}^{3/2}(q,c) - \frac{(4\alpha^2-1)}{16\alpha} I_{2n+1}^{5/2}(q,c) + \frac{\alpha}{4} I_{2n+2}^{5/2}(q,c) \right];$$

$$I_{13} = -4 \sum_{n=1}^{\infty} \int_0^{\infty} dy y e^{-y^2\left(\frac{1}{2}+\alpha\right)} \int_0^{\infty} dx x^4 e^{-2\alpha x^2} J_{2n}(xy) I_{2n+1}(2\alpha xy) =$$

$$= -\frac{1}{4\alpha^2} \sum_{n=1}^{\infty} \left[ (n+1) I_{2n}^{3/2}(q,c) + \frac{(4\alpha^2-1)}{16\alpha} I_{2n}^{5/2}(q,c) - \frac{(2n+3)}{4\alpha} I_{2n+1}^{3/2}(q,c) - \left( \frac{1}{4} + \frac{4\alpha^2-1}{32\alpha^2} \right) I_{2n+1}^{5/2}(q,c) + \right.$$

$$\left. + \frac{1}{8\alpha} I_{2n+2}^{5/2}(q,c) \right]; q = \frac{1+4\alpha+4\alpha^2}{8\alpha}, c = \frac{1}{2}, k = \frac{1}{\sqrt{2}} \left( 1 - \frac{q}{\sqrt{q^2+c^2}} \right)^{1/2}. \quad (A28)$$

In the case of the trial wave function $\varphi_0(x)$ such integrals do not appear. The contributions to the energy spectrum expressed by the integrals $I_1 - I_{13}$ are shown in the figure 6 in dependence on the parameter $\alpha$ of the trial wave function $\varphi_2(k)$.

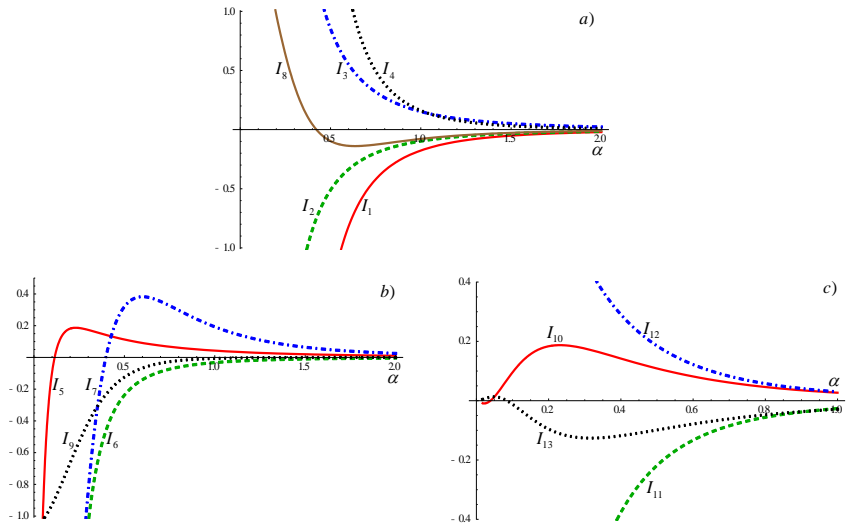

**Fig. 6.** The integrals $I_1 - I_{13}$ in dependence on the parameter $\alpha$ of the trial wave function $\varphi_2(k)$.